%% file: Nofrarias_LISASymp2012_DA.tex
\documentclass[11pt,twoside]{article}

\usepackage{asp2010}
\usepackage{graphicx}
\usepackage{color}

\resetcounters 

\begin{document}

\resetcounters

\title{State space modelling and data analysis exercises \\ in LISA Pathfinder}
\author{
M~Nofrarias$^{n}$, 
F~Antonucci$^{a}$,
M~Armano$^{b}$,
H~Audley$^{c}$,
G~Auger$^{d}$,
M~Benedetti$^{e}$,
P~Binetruy$^{d}$,
J~Bogenstahl$^{c}$,
D~Bortoluzzi$^{f}$, 
N~Brandt$^{g}$,
M~Caleno$^{h}$,
A~Cavalleri$^{a}$, 
G~Congedo$^{a}$,
M~Cruise$^{i}$,   
K~Danzmann$^{c}$, 
F~De~Marchi$^{a}$,
M~Diaz-Aguilo$^{j}$,
I~Diepholz$^{c}$, 
G~Dixon$^{i}$,
R~Dolesi$^{a}$, 
N~Dunbar$^{k}$, 
J~Fauste$^{b}$,
L~Ferraioli$^{d}$, 
V~Ferroni$^{a}$
W~Fichter$^{l}$, 
E~Fitzsimons$^{m}$,
M~Freschi$^{b}$,
C~Garc\'ia Marirrodriga$^{h}$,
R~Gerndt$^{g}$,
L~Gesa$^{n}$,
F~Gibert$^{n}$,
D~Giardini$^{o}$,
C~Grimani$^{p}$,
A~Grynagier$^{l}$,
F~Guzm\'an$^{c}$, 
I~Harrison$^{q}$,
G~Heinzel$^{c}$, 
M~Hewitson$^{c}$, 
D~Hollington$^{s}$, 
D~Hoyland$^{i}$,
M~Hueller$^{a}$,
J~Huesler$^{h}$,
O~Jennrich$^{h}$, 
P~Jetzer$^{t}$,
B~Johlander$^{h}$, 
N~Karnesis$^{n}$,
N~Korsakova$^{c}$,
C~Killow$^{m}$,
X~Llamas$^{u}$,
I~Lloro$^{n}$,
A~Lobo$^{n}$, 
R~Maarschalkerweerd$^{q}$,
S~Madden$^{h}$,
D~Mance$^{o}$, 
V~Martin$^{n}$,
I~Mateos$^{n}$, 
P~McNamara$^{h}$,
J~Mendes$^{q}$,
E~Mitchell$^{s}$,
D~Nicolodi$^{a}$, 
M~Perreur-Lloyd$^{m}$,
E~Plagnol$^{d}$, 
P~Prat$^{d}$,
J~Ramos-Castro$^{v}$,
J~Reiche$^{c}$,
J~A~Romera Perez$^{h}$,
D~Robertson$^{m}$, 
H~Rozemeijer$^{h}$,
G~Russano$^{a}$,
A~Schleicher$^{g}$, 
D~Shaul$^{s}$, 
C~F~Sopuerta$^{n}$,
T~J~Sumner$^{s}$, 
A~Taylor$^{m}$,
D~Texier$^{b}$,
C~Trenkel$^{k}$,
H~B~Tu$^{a}$,
S~Vitale$^{a}$, 
G~Wanner$^{c}$, 
H~Ward$^{m}$, 
S~Waschke$^{s}$,
P~Wass$^{s}$, 
D~Wealthy$^{k}$,
S~Wen$^{a}$,
W~Weber$^{a}$,
T~Ziegler$^{g}$,
P~Zweifel$^{o}$}

\affil{$^{a}$ Dipartimento di Fisica, Universit\`a di Trento and INFN,
Gruppo Collegato di Trento, 38050 Povo, Trento, Italy\\
$^{b}$ European Space Astronomy Centre, European Space Agency, Villanueva de la
Ca\~{n}ada, 28692 Madrid, Spain\\
$^{c}$ Albert-Einstein-Institut, Max-Planck-Institut f\"ur
Gravitationsphysik und Universit\"at Hannover, 30167 Hannover, Germany\\
$^{d}$ APC UMR7164, Universit\'e Paris Diderot, Paris, France\\
$^{e}$ Dipartimento di Ingegneria dei Materiali e Tecnologie
Industriali, Universit\`a di Trento and INFN, Gruppo Collegato di
Trento, Mesiano, Trento, Italy\\
$^{f}$ Dipartimento di Ingegneria Meccanica e Strutturale,
Universit\`a di Trento and INFN, Gruppo Collegato di Trento, Mesiano,
Trento, Italy\\
$^{g}$ Astrium GmbH, Claude-Dornier-Strasse, 88090 Immenstaad, Germany\\
$^{h}$ European Space Technology Centre, European Space Agency, 
Keplerlaan 1, 2200 AG Noordwijk, The Netherlands\\
$^{i}$ Department of Physics and Astronomy, University of
Birmingham, Birmingham, UK\\
$^{j}$ UPC/IEEC, EPSC, Esteve Terrades 5, 
E-08860 Castelldefels, Barcelona, Spain\\
$^{k}$ Astrium Ltd, Gunnels Wood Road, Stevenage, Hertfordshire, SG1 2AS, UK \\
$^{l}$ Institut f\"ur Flugmechanik und Flugregelung, 70569
Stuttgart, Germany\\
$^{m}$ School of Physics and Astronomy, University of
Glasgow, Glasgow, UK\\
$^{n}$ ICE-CSIC/IEEC, Facultat de Ci\`encies, 
E-08193 Bellaterra (Barcelona), Spain\\
$^{o}$ Institut f\"ur Geophysik, ETH Z\"urich, Sonneggstrasse 5, CH-8092, Z\"urich, Switzerland\\
}
\newpage
\author{
$^{p}$ Istituto di Fisica, Universit\`a degli Studi 
di Urbino/ INFN Urbino (PU), Italy
$^{q}$ European Space Operations Centre, European Space Agency, 64293 Darmstadt, Germany \\
$^{s}$ The Blackett Laboratory, Imperial College London, UK\\
$^{t}$ Institut f\"ur Theoretische Physik,
Universit\"at Z\"urich, Winterthurerstrasse 190, CH-8057 Z\"urich, Switzerland\\
$^{u}$ NTE-SENER, Can Mal\'e, E-08186, Lli\c{c}\`a d'Amunt,
Barcelona, Spain\\
$^{v}$ Universitat Polit\`ecnica de Catalunya, Enginyeria Electr\`onica,
Jordi Girona 1-3,  08034 Barcelona, Spain}

\begin{abstract} 
LISA Pathfinder is a mission planned by the European Space Agency (ESA) to test
the key technologies that will allow the detection of gravitational waves in 
space. 
The instrument on-board, the LISA Technology package, will undergo 
an exhaustive campaign of calibrations and noise characterisation campaigns
in order to fully describe the noise model. 
Data analysis plays an important role in the mission and for that reason 
the data analysis team has been developing a toolbox which contains
all the functionality required during operations. 
In this contribution we give an overview of recent activities, focusing
on the improvements in the modelling of the instrument and in the 
data analysis campaigns performed both with real and simulated data.
\end{abstract}

\input{include/Nofrarias_intro}
\input{include/Nofrarias_ltpda}
\input{include/Nofrarias_stoc}
\input{include/Nofrarias_realdata}
\input{include/Nofrarias_conclusions}
\input{include/Nofrarias_appendix}

\bibliography{Nofrarias_MyBibli}
\bibliographystyle{asp2010}

 \end{document}

%% file: include/Nofrarias_intro.tex
\section{Introduction}

LISA Pathfinder (LPF)~(ref. \cite{Antonucci12}) is the first step towards 
the detection of gravitational waves in space. 
The mission has been designed to test those key technologies 
required to detect gravitational radiation in space. 
In more concrete terms, the objective of this pioneering technology probe is to measure the differential acceleration between two free-falling test masses down to $3 \times 10^{-14} \rm m/s^2/\sqrt{\rm Hz}$
at 3\,mHz, with a measuring bandwidth 
from 1\,mHz to 0.1\,Hz. It is precisely in this low-frequency 
bandwidth where the observation of gravitational waves in space
can clearly contribute to our understanding of the Universe, 
since the gravitational wave sky is expected to be rich in 
interesting astrophysical sources in the millihertz band.

The main instrument on-board is the LISA Technology Package 
(LTP), which comprises subsystems which address the different functional requirements of the satellite: 
the Optical Metrology Subsystem (OMS)~(ref. \cite{Heinzel04}), 
the Gravitational Reference Sensor (GRS)~(ref. \cite{Dolesi03}), 
the Data and Diagnostics Subsystem 
(DDS)~(ref. \cite{Canizares11}) 
and the drag-free and attitude control system (DFACS)~(ref. \cite{Fichter05}). 
All of them working in closed loop in order to keep the two test 
masses on-board undisturbed at the required level.

The LPF mission has a planned duration of 200 days, 
during which a sequence of calibration and noise investigation runs are planned.
This compressed time schedule to achieve the scientific objectives was identified by the science team 
as having an important impact during mission operations, 
and for that reason a data analysis effort was started, which had to be parallel but in close contact with the hardware development.

In this contribution we give an overview of recent developments in the 
data analysis task. We will emphasise some recent developments, 
namely the usage of state space models to describe the LTP experiment and 
the exercises, both with real and simulated data, performed using the tools 
developed for the analysis of 
LISA Pathfinder.

%% file: include/Nofrarias_ltpda.tex
\section{The LTPDA toolbox}

During flight operations, the LTP experiment will undergo a tight schedule of experiments with the final aim 
of achieving the required differential noise acceleration between test masses.  
Some experiments could be strongly dependent on the results of previous experiments. 
For instance, noise models will need to be updated after a noise investigation to improve 
the description of the forthcoming investigations. 
Hence, data analysis will require a low latency between telemetry reception and 
the interpretation of the results. In order to cope with this demanding operational scenario it was decided to develop a data analysis framework specifically for use in LISA Pathfinder, the so-called \textsc{LTPDA} toolbox, a MATLAB$^{\copyright}$~(www.mathworks.com) toolbox which gathers together all 
analysis tools that will be used to analyse data during LISA Pathfinder operations. 
The main characteristics of this tool are the following:

\begin{itemize}

\item The toolbox is object-oriented. The user creates different types of \emph{objects}
to perform the analysis, the most usual ones are \emph{analysis objects} which act as
data containers, but there are others to allow many other functionalities like \emph{miir objects} to create infinite impulse response filters, 
\emph{pzmodel objects} to create pole-zero models, etc. 
More details can be found in~ref. \cite{Hewitson09}

\item Each object keeps track of the operations being applied to it. The object-oriented approach allows the adding of a \emph{history step} each time a method in the toolbox is applied to a given object.
That way, the object stores the history of operations being applied to it and, more importantly, this history can be used to re-run the entire analysis by any other user to cross-check the results.

\item During flight operations, the telemetry will be received at the
European Space Astronomy Center (ESAC) 
and then distributed to the different research institutions. 
The data analysis will be coordinated but spread between different centres around 
Europe. In order to ease the interchange of results, the toolbox provides the infrastructure to work through one or more database repositories.

\item The data analysis tools found to be relevant for the mission are implemented as methods in the toolbox.
The current list of methods cover topics like spectral estimators, parameter estimation methods, digital filters and transfer 
function modelling, time domain simulation, noise generation and data whitening, as well as
standard pre-processing tools like data splitting, interpolation, de-trending, and re-sampling.

\item The toolbox provides models for the different subsystems of LTP which the user can call
to perform simulations or to estimate parameters for a given data set. We give some more detail on the 
modeling in section \ref{sec.modeling}

\item As a part of the infrastructure of the toolbox, 
the development team has implemented a series of unit tests 
for each of the different methods which are executed on a daily basis to prevent the inclusion of 
new changes that may impact on the correct behaviour of the toolbox. 
About 6000 of these tests are executed repeatedly. Before any formal release to ESA\, more functional tests are run in order to check not only the 
individual blocks but also the toolbox as a whole.

\end{itemize}

A significant amount of effort has as well been put into the user manual in order to allow non-specialized 
users to implement their analysis using the toolbox. With the same purpose, the team has 
been running a series of training sessions, where users follow a guided tutorial to run 
standard analysis. These tutorials are available to any interested user in the toolbox's documentation.

\subsection{Modelling LISA Technology Package onboard LISA Pathfinder\label{sec.modeling}}

Modeling the instrument and the dynamics of the satellite is an important task among the data analysis activities. 
The LPF will be a complex experiment orbiting around 
L1 with hundreds of parameters determining 
its performance and many noise contributions to be determined. 
Disentangling the different contributions and 
dependencies will rely on our ability to build accurate models of the subsystems, 
which also need to be flexible enough to be able to include new information that the team may 
obtain during operations.

The team has been developing two modelling schemes. 
In chronological order, the first one uses the transfer function
of the system in the Laplace domain to encode the response of each subsystem. 
For instance, the interferometer measurement would be described as
 \begin{eqnarray}
  \vec{o}  & = &  ({\mathbf{M}}\cdot{\mathbf{S}}^{-1}+{\mathbf{C}})^{-1} (-
{\mathbf{C}}\,\vec{o}_i + \vec{g}_n + {\mathbf{D}}\cdot {\mathbf{S}}^{-1} \vec{o}_n),
\label{eq.dyn}
\end{eqnarray}
where $\mathbf{M}$ is the dynamical matrix, $\mathbf{C}$ is the
controller, and $\mathbf{S}$ stands for the sensing matrix, 
which translates the physical position of the test masses into
interferometer readout, $\vec{o}$. Subindex~$n$ stands for noise
quantities, either sensing noise ($\vec o_n$) or force noise ($\vec
g_n$) and subindex~$i$ stands for the injected signals ($\vec
o_i$). 
If we do not take into consideration the angular degrees of freedom, 
all of these will be 2-dimensional vectors with components referring to 
the  $\mathrm{x}_1$ (test mass \#1 displacement) and $\mathrm{x}_{\Delta}$ 
(differential displacement between test masses) channels respectively.
The matrices describing the motion of the test masses read as
\begin{eqnarray}
  \mathbf{M}  & = & \left(
  \begin{array}{cc}
  s^2 + \omega^2_{1}  + \frac{m_1}{m_{\rm SC}} \omega^2_1 + \frac{m_2}{m_{\rm SC}} \omega^2_2 & \frac{m_2}{m_{\rm SC}} \omega^2_2 \\ 
  \omega^2_{2} -\omega^2_{1} &  s^2 + \omega^2_{2}
  \end{array}   \right),    \\
  \mathbf{C} & = & \left(
  \begin{array}{cc}
   G_{\rm df} \, H_{\rm df} & 0\\
   0 &  G_{\rm sus} \, H_{\rm sus}
  \end{array}
  \right),  \label{eq.dyn}  \\
  \mathbf{S} & = & \left(
  \begin{array}{cc}
  S_{11} & S_{12}\\
   S_{21} & S_{22}
  \end{array} 
  \right), \label{eq.S}
\end{eqnarray}
where $\omega_{1}$ and $\omega_{2}$ are the stiffness,
coupling the motion of each test mass to the motion of the spacecraft;
$G_{\rm df}$ and $G_{\rm sus}$ are constant factors acting as
calibration factors of the controller, $H_{\rm df}(\omega)$ and $H_{\rm sus}(\omega)$. 
The latter encode the control laws of the loop;
the elements in the $\mathbf{S}$ matrix can be considered 
calibration factors and cross-couplings in the interferometer.
This type of model, based on transfer functions, have been shown to work successfully
when applied to the modelling of noise sources~(ref. \cite{Ferraioli11}) 
or in parameter estimation~(ref. \cite{Nofrarias10,Congedo12}).
\begin{figure}[!t]
\plotone{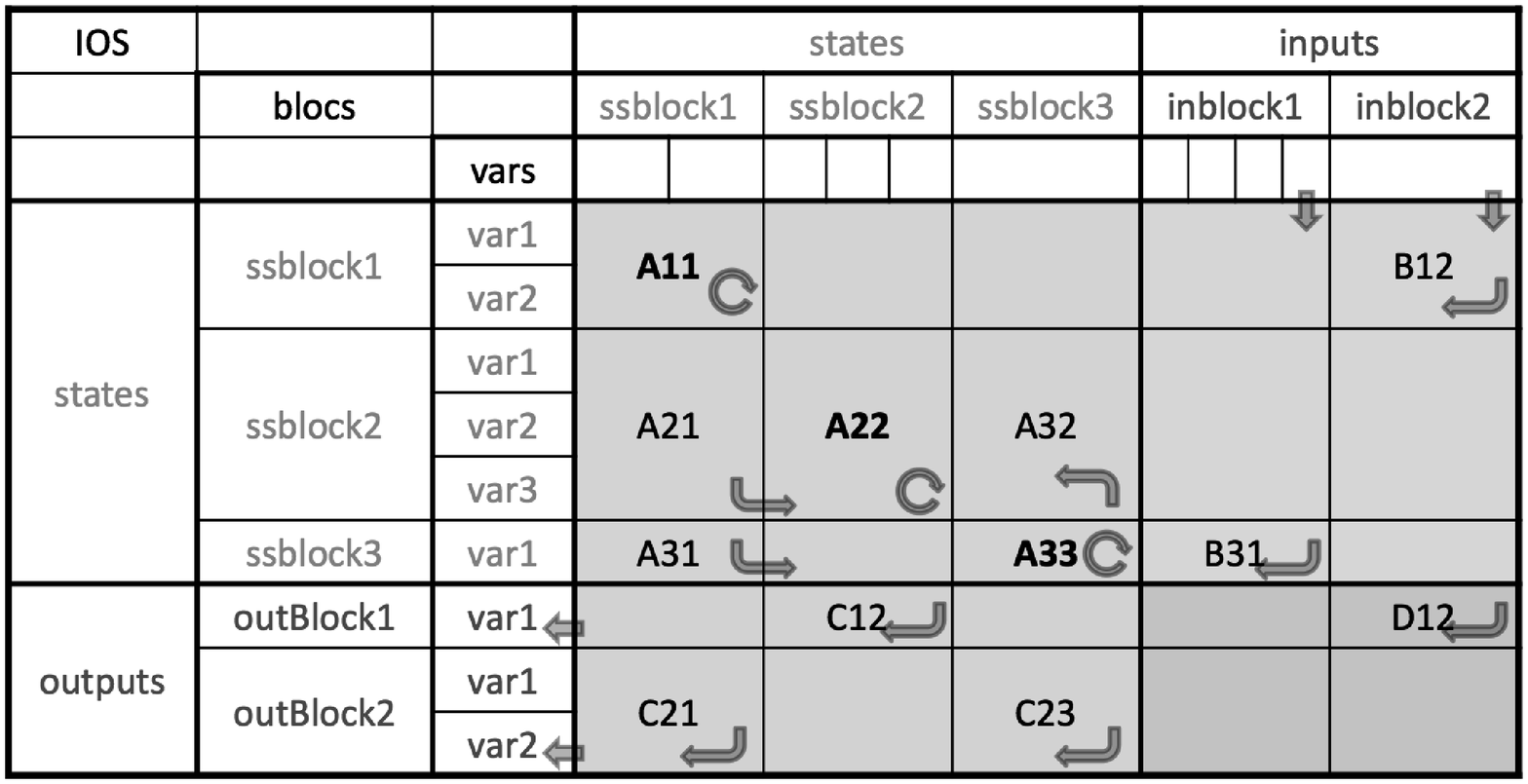}
\caption{A representation of a state space object in LTPDA. 
The ${\bf A}$, ${\bf B}$, ${\bf C}$ and ${\bf D}$ matrices are built in blocks. 
Each of these groups coefficients describing variables of the same nature. For 
instance, the diagonal block ${\bf A22}$ contains the dynamics of the variables 
of the states in block \#2, but the ${\bf A32}$ describes the impact of the 
dynamics of states in block \#3 on the states in block \#2.
\label{fig.ssmscheme} }
\end{figure}
\begin{figure}[!t]
\plotone{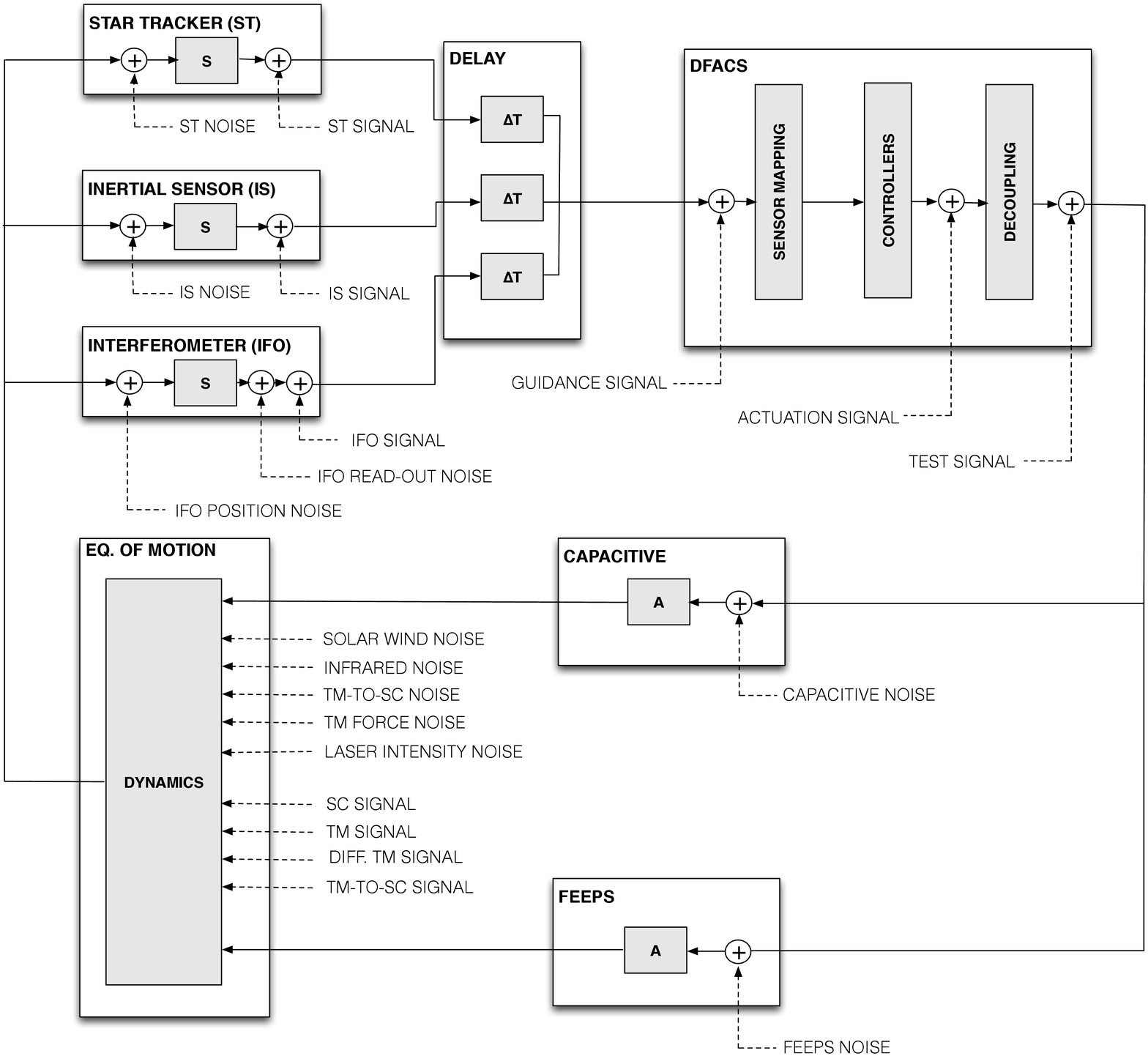}
\caption{A schematic representation of the LTP model.  
Each box is an independent state space model in LTPDA.
In the figure we also show the noise and signal injection ports, 
which allow to input a noise model or an injection signal 
to the state space model, respectively. \label{fig.ssmloop}}
\end{figure}

The second approach to model LPF is based on the state space representation~(ref. \cite{Kirk70}). 
This framework, widely used in control engineering, represents linear,
time invariant, dynamic systems as matrices of time-independent coefficients.
\begin{eqnarray}
\dot{\vec{x}}(t) & = & {\bf A} \cdot \vec{x}(t) + {\bf B} \cdot \vec{u}(t) \nonumber \\
\vec{y}(t) & = & {\bf C} \cdot \vec{x}(t) + {\bf D} \cdot \vec{u}(t) \label{eq.ssmdef}
\end{eqnarray}
where $\vec{x}(t)$ are the so called \emph{states}, i.e. variables describing the dynamical state of the system; 
$\vec{u}(t)$ represents the inputs and $\vec{y}(t)$ are the outputs of the system.
The matrices relating them encode the coefficients of the first-order differential equations:
$\bf A$ is the state matrix of the system, 
$\bf B$ is the input matrix, $\bf C$ is the output matrix 
and $\bf D$ is the feed-through matrix. 
This representation offers a fundamental advantage for the modelling of complex
instruments such as LISA Pathfinder, which is its modularity.  
By describing each subsystem in the LTP as in equation~(\ref{eq.ssmdef}),
we are able to build high-dimensionality systems, and at the same time simplify the process of model validation. 
A second advantage is that the state space description makes our modelling easier to scale:
the 1-dimensional models are built by selecting the relevant inputs, outputs and states
of the complete 3-dimensional version.

In our implementation~(ref. \cite{Grynagier08,AguilophD}), inputs, outputs and states are grouped into \emph{blocks}  with high level descriptions and global names. In that sense, the state space objects in LTPDA are block-defined, making easier to group together variables of similar nature --see Figure~\ref{fig.ssmscheme}.
The user is able to build models with multiple subsystems by \emph{assembling} two
state space models. For instance, one can choose to assemble a given 
model of the dynamics of the test masses with a model of the optical metrology subsystem or the  gravitational reference sensor. In such a case, the \emph{assemble}
method would look in those models for input and output blocks with the same name and link those variables (or \emph{ports}) that coincide in each of them.  

Figure \ref{fig.ssmloop} shows the main subsystems of LISA Pathfinder implemented 
as state space models, and how these are connected between them when working in closed-loop. 
As an example, in Appendix~\ref{sec.ifoExample} we show how to build a simplified state space model of the interferometer block. These blocks are already implemented in the LTPDA toolbox together with complete LPF built-in models, where all subsystems are already assembled in closed loop. Models are continuously being updated at the same time that new subsystems are added to our model library~(ref. \cite{Gibert12}).

%% file: include/Nofrarias_stoc.tex
\section{Testing the LTPDA toolbox with data}

Models and methods in LTPDA have been tested in several  exercises with real and simulated data. In the following we provide and overview of those, and give the
interested reader some references where to find more details.

\subsection{Simulated data: the operational exercises}

The first attempts to test the LTPDA tool against a real operations scenarion 
were organized as mock data challenges, following the concept previously applied
to LISA data analysis~(ref. \cite{Arnaud06}). In these, the team is split into data generation and 
data analysis in such a way that the latter faces the data analysis task without 
the complete knowledge of the parameters and settings defining the model used to generate the data, as it will happen during mission operations.
The first LPF mock data challenge dealt with a basic, 
albeit fundamental operation for the mission: the translation from displacement
to acceleration~(ref. \cite{ Ferraioli09, Monsky09}). In this first exercise, 
the model was agreed between data generation and data analysis groups. 
On the contrary, the second mock data challenge 
was based on an unknown model of the satellite for the data analysis and so 
parameter estimation techniques had to be applied~(ref. \cite{Nofrarias10}),
at the same time that the data generation techniques were improved~(ref. \cite{Ferraioli10}).

After these initial efforts, the following data analysis exercises
changed the objective of the activity: mock data challenge were substituted by operational 
exercises~( ref. \cite{Antonucci11_dataAnalysis}). The first relevant difference is that 
the OSE (Off-line Simulatons Environment), a detailed LISA Pathfinder non-linear simulator, is used to produce the data. Also, the aim of the operational exercises 
is to validate the experiments to be run on-board the satellite. 
In that sense, the operational exercises verify that the planned list of experiments are ready to be executed at the tele-command level. This list --put forward by the scientific team-- contains experiments to characterise the optical metrology~(ref. \cite{Audley11}) and the inertial sensor instrument~(ref. \cite{Dolesi03}); thermal~(ref. \cite{Canizares11}) and magnetic studies~(ref. \cite{Diaz-Aguilo12}), and also pure  free-fall experiments that aim at disentangling the contribution from actuators from the total noise budget~(ref. \cite{Grynagier09}). 
A series of parameter estimation studies have been performed using the 
OSE data, showing that our tools and models are able to explain and extract 
physical information, even if the model that we use for the analysis is not the same
as the one used to generate the data~(ref. \cite{Congedo12, Karnesis12, Nofrarias12}).
\begin{figure}[!t]
\plotone{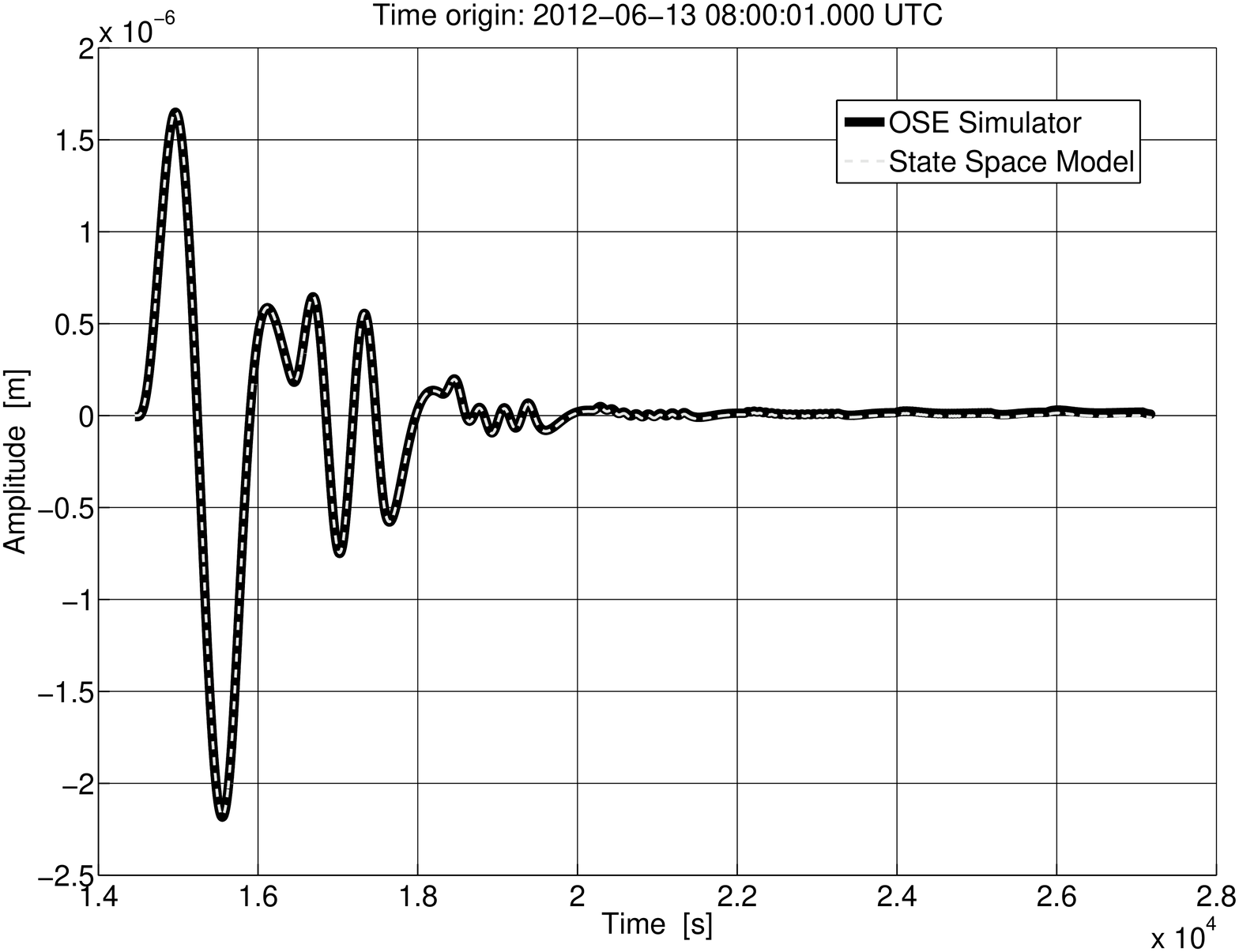}
\caption{Comparison between the OSE and state space model simulation. The same 
sinusoidal sequence is injected in the differential channel both in the 
OSE and in a LPF state space model, the injection point is the \emph{guidance signal} 
in Figure~\ref{fig.ssmloop}. In the same scheme, the response shown here would be the
output of the interferometer, after applying the delay block. 
\label{fig.stocssm}}
\end{figure}

Recently, the team has started a series of mission simulation campaigns focusing on 
the data analysis constraints during operations. 
In the first mission simulation the team was split in two different locations, 
ESAC (Madrid) and APC (Paris), to force a coordinated effort between different sites.  
During the three days duration of the exercise, data 
--previously generated with the OSE-- was received as in real operations: 
high priority (but low quality) telemetry in the morning and full telemetry at noon.
That way, the team simulated an operations scenario where
decisions based on the analysis results had to be taken on a daily basis.
In Figure~\ref{fig.stocssm} we show a comparison between the OSE and
the state space models obtained during this exercise. 
The LPF team plans to continue this activity increasing the degree of complexity 
in order to prepare for the real mission operations.

%% file: include/Nofrarias_realdata.tex
\subsection{Real data: testing the satellite on-ground}

Data analysis efforts have not only focused in simulated data. The tools and methods 
developed within the LTPDA toolbox are routinely used at the different laboratories
both to test the toolbox functionality and also to get the scientific team familiarized with it. 
There have been several testing campaigns, for instance~ref. \cite{Audley11}, 
where the toolbox has been used to analyse data from the real hardware.  

As the launch draws nearer these test campaigns become more realistic, involving
more subsystems assembled in the spacecraft. One of these was 
the space-craft closed loops test, performed 
at Astrium Ltd. premises. The focus here
was to validate the ability of the space-craft to perform the science goals. However,
the main instruments inside, like the optical metrology or the inertial sensor, where 
substituted by equipments simulating their operations. Communications and telemetry 
from the satellite platform where as in-flight operations and the whole processing chain 
underwent a realistic scenario. 

A second testing campaign testing the real hardware was performed
in the space simulator at IABG facilities. The so called On-Station Thermal Test (OSTT)
was a twofold objective campaign, testing the operations of the optical metrology subsystem in a realistic space environment at the same time that the space-craft was
subjected to a thermal balance test at two extreme temperature levels. 
More details on this campaign can be found in the 
following contribution on this volume~ ref. \cite{Guzman12}.

%% file: include/Nofrarias_conclusions.tex
\section{Summary}

Given the complexity and the tight schedule, the in-flight operations of LISA Pathfinder will require a well coordinated and efficient data analysis, ready to
react with a short time schedule once the data are received. 
Tools being developed in the LTPDA toolbox for that purpose are already in a 
mature state and the experiments to be performed on the satellite have been simulated and 
analysed with these methods.

In this contribution we have provided an overview of two of the many aspects involved in the 
data analysis of the LISA Pathfinder mission, i.e. the modelling and the data analysis exercises 
being performed to test the data analysis tools. 
On the modelling side, an important effort has been put forward to develop a framework 
that allows an efficient and flexible scheme to build models that will be afterwards 
used to understand the data from the satellite. 
These are currently implemented as state space models, 
which proved also to add advantages in terms of infrastructure maintenance.

The implemented data analysis tools have been tested with simulated data, 
mimicking the characteristics of the LISA Pathfinder data.
Initially, these exercises focused on the development of data analysis algorithms.
However, as the algorithms reached a mature state, the focus of the exercises 
moved to apply those algorithms to the experiments to be performed
in flight. This is the main motivation behind what we call operational exercises. 
In parallel to these activities, the LPF mission has entered a phase of on-ground testing, 
which will be a perfect opportunity for the data analysis team to deal with real satellite
data. Analysis and further results on these are currently ongoing and will be reported 
elsewhere.

%% file: include/Nofrarias_appendix.tex
\section*{Appendix}

\appendix

\section{A modelling example: a perfect interferometer model \label{sec.ifoExample}}

Let's assume that we are interested in modelling a perfect interferometer. This is, by itself, a complex instrument~(ref. \cite{Heinzel04}) but here we will be interested in the contribution of the interferometer to the dynamics of the satellite. In that sense, the interferometer must be understood as a sensor that translates the physical displacement of the test mass into a phase measurement from which an on-board algorithm calculates the actual attitude of the test mass which will be taken over by the DFACS controller as indicated in Figure~\ref{fig.ssmloop}.

In the transfer function description this information is captured in the {\bf S} matrix. 
Hence, for an interferometer that does not add any cross-coupling this would turn into the identity matrix,
\begin{eqnarray}
  \mathbf{S} & = & \left(
  \begin{array}{cc}
  1 & 0\\
  0 & 1
  \end{array} 
  \right)
\end{eqnarray}

On the other hand, in the state space model case, the system will be fully defined by the four matrices in equation~(\ref{eq.ssmdef}). As in the previous case, we model the interferometer purely as a sensor and hence, it translates the input (the displacement of the test masses) to the interferometer output. The only matrix with non-zero coefficients 
is the $\bf D$~matrix, there is no contribution to the dynamics 
of the states by themselves ($\bf A$~matrix), neither coming from the inputs to the system ($\bf B$~matrix) or any dependence of the output with respect the states of 
the system ($\bf C$~matrix).  A perfect interferometer model would be given by 

\begin{eqnarray}
  \mathbf{A} & = & \left(
  \begin{array}{cc}
  0 & 0\\
  0 & 0
  \end{array} 
  \right), \quad
   \mathbf{B}  =  \left(
  \begin{array}{cc}
  0 & 0\\
  0 & 0
  \end{array} 
  \right), \quad
\mathbf{C}  =  \left(
  \begin{array}{cc}
  0 & 0\\
  0 & 0
  \end{array} 
  \right), \\
   \mathbf{D} & = & \left(
  \begin{array}{cccc}
  \mathbf{D_{11}} & \mathbf{D_{12}} & \mathbf{D_{13}} & \mathbf{D_{14}}\\
  \end{array} 
  \right)
  \end{eqnarray}
where the $\mathbf{D}$ matrix shows the block structure of our implementation that we have 
previously described. These matrices read as
\begin{eqnarray}
  \mathbf{D_{11}} & = & \left(
  \begin{array}{cc}
  1 & 0\\
  -1 & 1
  \end{array} 
  \right), \quad
   \mathbf{D_{12}}  =  \left(
  \begin{array}{cc}
  1 & 0\\
  0 & 1
  \end{array} 
  \right), \\
\mathbf{D_{13}}  & = &   \left(
  \begin{array}{cc}
  1 & 0\\
  -1 & 1
  \end{array} 
  \right), \quad
   \mathbf{D_{14}}  =  \left(
  \begin{array}{cc}
  1 & 0\\
  0 & 1  \end{array} 
  \right)
  \end{eqnarray}
and correspond to the four input blocks of the interferometer in our implementation, which will be, in corresponding order with the previous matrices
\begin{eqnarray}
  \vec{x} & = & \left(
  \begin{array}{c}
  x_1 \\
  x_2
  \end{array} 
  \right), \quad
   \vec{n_R}  =  \left(
  \begin{array}{cc}
  n_{R1} \\
  n_{R2}
  \end{array} 
  \right), \\
  \vec{n_P}  & = &   \left(
  \begin{array}{cc}
  n_{P1} \\
  n_{P2}
  \end{array} 
  \right), \quad
   \vec{s}  =  \left(
  \begin{array}{cc}
  s_1 \\
  s_2
 \end{array} 
  \right)
  \end{eqnarray}
which are: the test masses physical displacements ($\vec{x}$), the read-out interferometer noise ($\vec{n_{R}}$), the test mass position noise ($\vec{n_P}$), and an input block to inject signals in the interferometer ($\vec{s}$). The output of the interferometer will be the displacement of the test masses as translated in phase shifts. The output block, expressed as a vector,  would be  
\begin{eqnarray}
  \vec{o} & = & \left(
  \begin{array}{c}
  o_1 \\
  o_{\Delta}
  \end{array} 
  \right)
\end{eqnarray}
where we notice that, opposite to the transfer function case, 
in the state space modelling the differential channel is built in the interferometer since the
inputs to the interferometer were the displacements of the two test masses, $x_1$ and $x_2$.